\documentclass[aps,prd,superscriptaddress,showkeys,showpacs,twocolumn,a4paper]{revtex4-1}
\usepackage{ulem}
\usepackage{color}
\usepackage{graphicx}
\usepackage{appendix}
\usepackage{epsfig}
\usepackage[colorlinks=true,linkcolor=red]{hyperref}
\usepackage{epstopdf}
\usepackage{amsmath}
\usepackage{subfig}
\usepackage{comment}
\usepackage{float}
\usepackage[dvipsnames]{xcolor}
\newcommand{\be}{\begin{equation}}
\newcommand{\ee}{\end{equation}}
\newcommand{\ba}{\begin{eqnarray}}
\newcommand{\ea}{\end{eqnarray}}

\begin{document}

\title{Heavy quark transport in an anisotropic hot QCD medium: Collisional and Radiative processes}

\author{Jai Prakash}
\email{jai183212001@iitgoa.ac.in}
\affiliation{School of Physical Science, Indian Institute of Technology Goa, Ponda-403401, Goa, India}

\author{Manu Kurian}
\email{manu.kurian@iitgn.ac.in}
\affiliation{Indian Institute of Technology Gandhinagar, Gandhinagar-382355, Gujarat, India}

\author{Santosh K. Das}
\email{santosh@iitgoa.ac.in}
\affiliation{School of Physical Science, Indian Institute of Technology Goa, Ponda-403401, Goa, India}

\author{Vinod Chandra}
\email{vchandra@iitgn.ac.in}
\affiliation{Indian Institute of Technology Gandhinagar, Gandhinagar-382355, Gujarat, India}


\begin{abstract}
The impact of momentum anisotropy on the heavy quark transport coefficients due to collisional and radiative processes in the QCD medium has been studied within the ambit of kinetic theory. Anisotropic aspects (momentum) are incorporated into the heavy quark dynamics through the non-equilibrium momentum distribution function of quarks, antiquarks, and gluons.  These non-equilibrium distribution functions that encode the physics of momentum anisotropy and turbulent chromo-fields have been obtained by solving the ensemble-averaged diffusive Vlasov-Boltzmann equation.  The momentum dependence of heavy quark transport coefficients in the medium is seen to be sensitive to the strength of the anisotropy for both collisional and radiative processes. In addition, the collisional and radiative energy loss of the heavy quark in the anisotropic hot QCD medium have been analyzed. The effects of anisotropy  on the drag and diffusion coefficients are observed to have a visible impact on the nuclear suppression factor both at the RHIC and LHC.
\end{abstract}



\maketitle

 \section{Introduction}
The vibrant experimental programs on heavy-ion collisions pursued at the Relativistic Heavy Ion Collider (RHIC) and the Large Hadron Collider (LHC) have strongly suggested the existence of hot and dense phase of nuclear matter, known as the quark-gluon plasma (QGP)~\cite{Adams:2005dq, Adcox:2004mh,Back:2004je,Arsene:2004fa,Aamodt:2010pb}. The space-time evolution of the QGP medium is successfully studied within the framework of relativistic viscous hydrodynamics~\cite{Gale:2013da, Heinz:2013th, Jaiswal:2016hex}. Experimental observations along with the hydrodynamical description suggest that the QGP behaves like near-perfect with a tiny value for the shear viscosity to entropy ratio $\eta/s$ (except for the regions very close to the transition temperature $T_c$ where the bulk viscosity to entropy ratio $\zeta/s$ may have a larger value) and most vortical fluid~\cite{STAR:2017ckg,Romatschke:2007mq}. The impact of non-zero $\zeta/s$ in the QGP evolution has also been explored recently ~\cite{Ryu:2015vwa}.

Among reliable signatures from experimental observations, heavy quarks, especially charm and bottom quarks, are identified as the effective probes to study the properties of the QGP~\cite{Rapp:2018qla,Dong:2019unq,Prino:2016cni,Aarts:2016hap,Andronic:2015wma}. Heavy quarks are mostly created in the initial stages of the collisions and undergo Brownian motion in the QGP medium owing to its large mass in comparison with the temperature of the medium, $i.e.$, $M_{HQ}>>T$ (where $M_{HQ}$ and $T$ denote the heavy quark mass and temperature of the thermalized medium, respectively). Heavy quarks witness the expansion of the created hot fireball and can carry information about the hotter phases of the matter and the initial conditions of the heavy-ion collisions. While traveling through the medium, heavy quarks loose their energy due to the interactions with the constituent medium particles, and its dynamics can be described within the Fokker-Planck approach~\cite{Svetitsky:1987gq,GolamMustafa:1997id}. The interactions of the heavy quarks with the medium particles are embedded through the drag and diffusion coefficients of the heavy quark.  Hence, the study of heavy quark transport coefficients in the QGP medium is a field of high contemporary interest, see Refs.~\cite{Cao:2018ews,Xu:2018gux,Kurian:2020orp,Das:2013kea,Banerjee:2011ra} for the recent studies. The two dominant processes that contribute to the energy loss of heavy quark in the medium are namely, the collisional (elastic interaction) and the inelastic interactions like gluon bremsstrahlung~\cite{Mustafa:2004dr,Braaten:1991we,Sarkar:2018erq,Abir:2012pu,Liu:2020dlt,Cao:2013ita,Zigic:2018ovr}. The heavy quark transport, its energy loss, and the associated physical observables such as nuclear suppression factor $R_{AA}$, flow coefficients, etc have been investigated in several works in the literatures~\cite{vanHees:2005wb,Gossiaux:2008jv,Cao:2013ita,Das:2015ana,Scardina:2017ipo,Cao:2016gvr,Song:2015sfa,Adare:2006nq,Adler:2005xv,Alberico:2013bza,vanHees:2007me,Li:2019wri,Jamal:2021btg,Akamatsu:2008ge,Uphoff:2011ad,Xu:2013uza,Zigic:2018ovr}. Recently, drag and diffusion coefficients of the heavy quark undergoing radiative loss by soft gluons emission along with the collisional interactions have been studied in Ref.~\cite{Mazumder:2013oaa}. However, many estimations considered the QGP as a thermalized static medium. It is an interesting direction to investigate the impact of momentum anisotropy of the medium on the radiative and collisional processes of heavy quarks as it may reflect on the final stage observables.

The effect of non-equilibrium dynamics of the QGP has already been explored in photon production, dilepton emission, heavy quarkonia, and other associated physical observables at the RHIC and LHC~\cite{Shen:2014nfa,Dusling:2008xj,Vujanovic:2013jpa,Thakur:2020ifi,Chandra:2020hik,Chandra:2016dwy}. Recently, some works have been done to analyse the effects of shear and bulk viscous coefficients on heavy quark transport in the expanding medium~\cite{Kurian:2020orp,Song:2019cqz,Das:2012ck,Singh:2019cwi,Kurian:2020kct}.

The momentum anisotropy present in the QGP medium  may induce instability to the Yang-Mills fields (termed as Chromo-Weibel instability). Studies have shown that the physics of the Chromo-Weibel instability may have a significant role in understanding the properties of the QGP and its evolution in heavy-ion collision experiments~\cite{Mrowczynski:1993qm,Randrup:2003cw,Mrowczynski:1988dz,Romatschke:2003ms}. Such instabilities to the Yang-Mills field equations in the rapidly expanding QGP may lead to plasma turbulence as described in Ref.~\cite{Asakawa:2006tc}. In Refs.~\cite{Asakawa:2006tc,Majumder:2007zh,Asakawa:2006jn}, the authors have realized that the instability while coupled with the rapid expansion of the medium leads to the  anomalous transport in the medium in the similar lines as has been argued long back by Dupree~\cite{A} in the case of QED plasma. The anomalous transport processes due to the turbulent fields can be explored by analyzing various signals emitted during the fireball expansion and may provide a possible explanation of the nearly perfect liquidity of the QCD medium created in the heavy-ion collisions~\cite{Asakawa:2006tc,Majumder:2007zh}. The physics of anisotropy and related aspects of anomalous transport could be captured in the non(near)-equilibrium distribution functions of quarks, antiquarks and gluons~\cite{Chandra:2016dwy}. In Ref.~\cite{Chandra:2015gma}, the impact of Chromo-Weibel instability on the heavy quark transport coefficients for the collisional process in the hot  QCD medium has been investigated.  Recently performed non-equilibrium calculations have shown significant dependence on momentum broadening, energy loss, and the associated observables in the heavy-ion collisions, see the Refs.~\cite{Das:2017dsh,Mrowczynski:2017kso,Ruggieri:2018rzi,Boguslavski:2020tqz,Carrington:2020sww,Sun:2019fud} for details. The current focus is to explore the heavy quark dynamics in the anisotropic (momentum)  hot  QCD medium while considering both the collisional and radiative processes of heavy quarks in the medium by incorporating the effects of anisotropy along the lines of Ref.~\cite{Chandra:2015gma}. This is  perhaps the first attempt where the physics of momentum anisotropy leading to anomalous transport processes has been incorporated in the heavy-quark dynamics.  The anisotropic aspects of the QGP medium have been observed to play a prominent role in the heavy quark transport coefficients, its energy loss and the associated Nuclear suppression factor $R_{AA}$.

The manuscript is organized as follows. Section II is devoted to the description of near-equilibrium distribution functions of quarks and gluons in the anisotropic medium. The mathematical formulation of the heavy quark transport in the medium while considering the collisional and radiative processes are discussed in section III. The results and the followed discussions are presented in Section IV. Finally, in section V, we summarize the present analysis with an outlook.\\
  
\noindent {\bf Notations and conventions}: The subscript $k$ in the analysis represents the particle species, $i.e.$, $k=(g, \Tilde{q})$, where $\Tilde{q}$ and $g$ denotes quarks and gluons, respectively. The quantity $g_{\mu\nu}$ is the metric tensor and $u^{\mu}$ represents the normalized fluid four-velocity such that $u^{\mu}u_{\mu}=1$. The traceless symmetric velocity gradient can be defined as $\Delta u_{\mu\nu}=\frac{1}{2}(\nabla_\mu u_\nu+\nabla_\nu u_\mu)-\frac{1}{3}g_{\mu\nu}\nabla_\gamma u^\gamma$. We have $u=\frac{z}{\tau}$ and $\Delta u_{ij}=\frac{1}{3}\text{diag}(-1, -1, 2)$, where $\tau$ denotes the proper time of expansion, for the boost invariant $1+1-$D Bjorken's flow.
\section{Momentum distributions of quarks and gluons in anisotropic medium}
An adequate modeling of the momentum distribution functions of quarks and gluons are necessary to encode the thermal medium effects in the analysis of heavy quark transport in the QGP. The realistic equation of state (EoS) effects are incorporated in the EQPM description of the QCD medium via effective fugacities of quarks and gluons, $z_{\Tilde{q}}$ and $z_g$ respectively~\cite{Chandra:2011en,Chandra:2008kz}. In equilibrium, the EQPM momentum distribution function has the form,
\begin{equation}\label{1}
f^0_{k}=\frac{z_{k}\exp{(-\beta E_{q})}}{1-a_k z_{k}\exp{(-\beta E_{q})}},
\end{equation} 
with $a_g=1$ for gluons and $a_q=-1$ for quarks, and $E_{q}=|{\bf q}| \equiv q$ for quarks (massless limit) and gluons.  The fugacity parameter modifies the single particle dispersion relation as,
\begin{align}\label{2}
&\omega_{k}=E_q+\delta\omega_k,  &&\delta\omega_k=T^{2}\partial_{T} \ln(z_{k}).
\end{align}
The term $\delta\omega_k$ is the modified part of the energy dispersion and is related to the quasiparticle collective excitation in the medium. The temperature dependence of the quark and gluon fugacity parameters can be described from realistic $(2+1)$ flavor lattice QCD EoS~\cite{Borsanyi:2013bia}. To obtain the nonequilibrium distribution functions of quarks and gluons in a rapidly expanding medium with an anisotropy, one needs to solve the Vlasov–Boltzmann equation in the presence of turbulent color fields. To that end, we consider the near-equilibrium distribution function as,
\begin{align}\label{3}
&f_{k}({{\bf q}},{{\bf r}}) = f^0_{k}+\delta f_k, &&\delta f_k=(1+a_k f^0_{k})f^1_{k}({{\bf q}}),
\end{align}
where $f_k^1$ is the linear perturbation to the distribution function of the $k-$th species. We employ the following ansatz for the linear perturbation for the quasiparticles, 
\begin{align}\label{4}
f^1_{k}({{\bf q}}) = -\frac{\bar{\Delta}_{k}({{\bf q}})}{\omega_{k} T^2 \tau}\left(q_{z}^2 - \frac{q^2}{3} \right),
\end{align}
where the quantity $\bar{\Delta}({{\bf q}})$ denotes the strength of momentum anisotropy in the QGP medium. In general, the evolution of the particle momentum distribution function in the medium can be described by setting up the Vlasov–Boltzmann equation as follows~\cite{Heinz:1983nx},
\begin{align}
v^{\mu}\dfrac{\partial}{\partial x^{\mu}}{f}({\bf r}, {\bf q}, t)+g {\bf F}^a.{\mathbf{\nabla}}_qf^a({\bf r}, {\bf q}, t)=0,
\end{align}
where ${\bf F}^a={\bf E}^a+({\bf v}\times{\bf B}^a )$ is the color Lorentz force. The quasiparticle distribution and the color-octet distribution function $f^a({\bf r}, {\bf q}, t)$ can be defined as the moments of the distribution function $\Tilde{f}^a({\bf r}, {\bf q}, \mathcal{Q}, t)$ in an extended phase space that includes the color sector as follows,
\begin{align}
   &{f}({\bf r}, {\bf q}, t)=\int d\mathcal{Q} \Tilde{f}^a({\bf r}, {\bf q}, \mathcal{Q}, t),\\
   &{f}^a({\bf r}, {\bf q}, t)=\int d\mathcal{Q} \mathcal{Q}^a\Tilde{f}^a({\bf r}, {\bf q}, \mathcal{Q}, t),
\end{align}
where $\mathcal{Q}$ is the color charge. In the anisotropic QGP medium, the color field is turbulent, and its action on the quasiparticles can be described by taking an ensemble average. This analysis has been initially done in Refs.~\cite{Asakawa:2006tc,Asakawa:2006jn} for the ultra-relativistic gas of quarks/antiquarks and gluons, and later extended to the interacting QGP within the EQPM~\cite{Chandra:2015gma}. Following the same formalism, the ensemble-averaged by the diffusive Vlasov-Boltzmann equation can be described as follows,
\begin{equation}\label{5}
v^{\mu}\dfrac{\partial}{\partial x^{\mu}}\bar{f}-\mathcal{F}_A\bar{f}=0,
\end{equation}
where $\bar{f}$ represents the ensemble-averaged distribution of the particles and in the current analysis, we have $\bar{f}\equiv f_{k}$. The force term can be defined in terms of color averaged chromo-electromagnetic fields as,
\begin{align}\label{6}
 \mathcal{F}_A\bar{f}=&-\dfrac{g^2C_2}{3(N_c^2-1)\omega^2_{k}}\langle E^2+B^2\rangle_{k}\tau_m\nonumber\\
&\times \mathcal{L}^2f^0_{k}(1+a_k f^0_{k})q_iq_j\Delta u_{ij},
\end{align}
where $\tau_m$ quantifies the time scale of instability in the QGP medium and $C_2$ is the Casimir invariant $SU(N_c)$ theory. Here, the operator $\mathcal{L}^2$ takes the form as follows,
\begin{equation}\label{7}
 \mathcal{L}^2=\mid{\bf q}\times\partial_{{\bf q}}\mid^2-\mid{\bf q}\times\partial_{{\bf q}}\mid^2_z.
\end{equation} 
The contribution to the system dynamics from the leading order collisional processes can be quantified in terms of collision kernel in the transport equation. Note that the focus of the current study is on the anomalous contributions to the quasiparticle momentum distribution functions. To that end, we will not consider the collisional effects within the bulk medium in the analysis. This assumption is also based on the fact that anomalous transport is the dominant mechanism and leads to significant suppression of the transport coefficients in the expanding medium~\cite{Chandra:2020hik}. We intend to work on the interplay of collisional and anomalous processes in the QGP medium in the near future. Employing the form of equilibrium quasiparticle distribution function as defined in Eq.~(\ref{1}) and following the same formalism in Ref.~\cite{Asakawa:2006tc}, one can obtain the form of $\bar{\Delta}({{\bf q}})$ by solving the Boltzmann equation as,
\begin{equation}\label{8}
\bar{\Delta}({{\bf q}})= 2(N_c^2-1)\dfrac{\omega_{k}T}{3g^2C_2\langle E^2+B^2\rangle_{k}\tau_m}.
\end{equation} 
The unknown factors in the denominator of Eq.~(\ref{8}) can be related to the jet quenching parameter $\hat{q}$ in both quark and gluonic sectors~\cite{Majumder:2007zh}. Shear viscosity and jet quenching parameter are the two crucial coefficients that may get a significant impact from the turbulent fields. In Ref.~\cite{Asakawa:2010xf}, the authors have realized that the parameter $\hat{q}$ is proportional to the mean momentum square per unit length on the particle imparted by turbulent color fields. The unknown factor $\langle E^2+B^2\rangle_{k}\tau_m$ can be related to the jet quenching parameter as~\cite{Majumder:2007zh,Chandra:2011bu},
\begin{equation}\label{9}
\hat{q}=\dfrac{2g^2C_{g/f}}{3(N_c^2-1)}\langle E^2+B^2\rangle\tau_m,
\end{equation}
where $C_g=N_C$ for gluons and $C_f=\frac{N^2_C-1}{2N_C}$ for quark sector. Substituting Eq.~(\ref{8}) and Eq.~(\ref{9}) in Eq.~(\ref{4}), we obtain the near-equilibrium distribution function as,
\begin{equation}\label{10}
  f_{k}({{\bf q}},{{\bf r}}) = f^0_{k}-(1+a_k f^0_{k})\frac{4\omega_k}{9\hat{q}_kT \tau}\Big(q_{z}^2 - \frac{q^2}{3} \Big).
\end{equation}
Let us now proceed to the investigation of the heavy quark drag and diffusion coefficients due to the collisional and radiative processes in the anisotropic QGP medium.
\section{Formalism: Heavy quark drag and momentum diffusion}
In the present analysis, we adopt the formalism developed by Svetitsky~\cite{Svetitsky:1987gq} such that the evolution of heavy quark in the medium can be considered as Brownian motion. The dynamics of heavy quark can be described in terms of the  distribution function within the framework of transport theory as,
\begin{equation}\label{11}
  p^{\mu}\partial_{\mu}f_{HQ}=\bigg(\dfrac{\partial f_{HQ}}{\partial t}\bigg)_{\text{int}},
\end{equation}
where $f_{HQ}$ is the heavy quark momentum distribution. The elastic and inelastic processes of the heavy quark in the medium modify the distribution function, and the rate of change of $f_{HQ}$ due to the interactions can be quantified in terms of the collision term $\Big(\frac{\partial f_{HQ}}{\partial t}\Big)_{\text{int}}$ as follows,
\begin{align}\label{12}
   \bigg(\dfrac{\partial f_{HQ}}{\partial t}\bigg)_{int}=&\int{d^3{\bf k}\bigg[\omega({\bf p}+{\bf k},{\bf k})f_{HQ}({\bf p}+{\bf k})}\nonumber\\
   &-\omega({\bf p},{\bf k})f_{HQ}({\bf p})\bigg],
\end{align}
where $w({\bf p,k})$ denotes the rate of collision for heavy quarks with the constituent particles in the medium such that its momentum changes from {\bf p} to {\bf p}-{\bf k}. Owing to the large mass of heavy quark, the Boltzmann equation can be simplified by considering the Landau approximation, $i.e.$, the momentum transfer of the heavy quark is soft (${\bf |p|}\gg {\bf |k|}$). Now, expending $w({\bf p+k,k})f({\bf p,k})$ up to second order of ${\bf k}$, we have 
\begin{align}\label{13}
   \omega({\bf p+k,k})f_{HQ}({\bf p,k})\approx&\,\omega({\bf p,k})f_{HQ}({\bf p})+{\bf k}.\frac{\partial}{\partial {\bf p}}(\omega f_{HQ})\nonumber\\ &+\frac{1}{2}k_ik_j\frac{\partial^2}{\partial p_i\partial p_j}(\omega f_{HQ}).
\end{align}
Incorporating Eq.~(\ref{13}) in Eq.~(\ref{12}), the relativistic non-linear transport equation reduces to the Fokker-Planck equation as follows,
\begin{align}\label{14}
  	\frac{\partial f_{HQ}}{\partial t}=\frac{\partial}{\partial p_i}\left[A_i({\bf p})f_{HQ}+\frac{\partial}{\partial p_j}\Big(B_{i j}({\bf p})f_{HQ}\Big)\right],
  	\end{align}
where $A_i$ and $B_{ij}$ respectively quantify the drag force and momentum diffusion of the heavy quarks in the medium due to the interactions.
\subsection{ { \bf{ Collisional process}}}
For the elastic two-body collisional process $HQ(P)+l(Q)\rightarrow HQ(P^{'})+l(Q^{'})$, where $l$ represents constituent particles in the medium (quarks, antiquarks, and gluons) and $P, Q$ are the four-momentum of heavy quark and constituent particle before the collision, the heavy quark drag and momentum diffusion can be described as, 
 \begin{align}\label{15}
    A_i=&\frac{1}{2E_p}\int{\frac{d^3{\bf q}}{(2\pi)^32E_q}}\int{\frac{d^3{\bf q}'}{(2\pi)^32E_{q'}}}\int{\frac{d^3{\bf p}'}{(2\pi)^32E_{p'}}}\frac{1}{\gamma_{HQ}}\nonumber\\ &\times\sum|\mathcal{M}_{2\rightarrow 2}|^2(2\pi)^4\delta^4 (P+Q-P'-Q') f_{k}({\bf{q}})\nonumber\\ &\times\Big(1+a_k f_{k}({\bf{q'}})\Big)\Big[({\bf p}-{\bf p}')_i\Big]\nonumber\\
	&=\langle\langle({\bf p}-{\bf p}')_i\rangle\rangle,
\end{align}
  and 
\begin{align}\label{16}
    B_{ij}=&\frac{1}{2E_p}\int{\frac{d^3{\bf q}}{(2\pi)^32E_q}}\int{\frac{d^3{\bf q}'}{(2\pi)^32E_{q'}}}\int{\frac{d^3{\bf p}'}{(2\pi)^32E_{p'}}}\frac{1}{\gamma_{HQ}}\nonumber\\ &\times\sum|\mathcal{M}_{2\rightarrow 2}|^2(2\pi)^4\delta^4 (P+Q-P'-Q') f_{k}({\bf{q}})\nonumber\\ &\times\Big(1+a_k f_{k}({\bf{q'}})\Big)\frac{1}{2}\Big[({\bf p}-{\bf p}')_i({\bf p}-{\bf p}')_j\Big]\nonumber\\
	&=\frac{1}{2}\langle\langle({\bf p}-{\bf p}')_i({\bf p}-{\bf p}')_j\rangle\rangle,
\end{align} 
respectively. Note that the delta function ensures the energy-momentum conservation and $f_{k}$ is the near-equilibrium phase space distribution for the light quarks and gluons as described in  Eq.~(\ref{3}). Here, $\gamma_{HQ}$ denotes the statistical degeneracy factor of the heavy quark and   $|\mathcal{M}_{2\rightarrow 2}|$ represents the matrix element for the two-body elastic collisions of heavy quarks with light quarks, antiquarks, and gluons~\cite{Svetitsky:1987gq}. It is important to emphasize that the heavy quark drag quantifies the thermal average of the momentum transfer, whereas the momentum diffusion measures the square of the momentum transfer due to the interaction. As $A_{i}$ depends on the momentum, we have the following decomposition the heavy quark drag,
\begin{align}\label{17}
&A_{i}=p_{i}A(p^2), &&A=\langle\langle 1 \rangle\rangle - \frac{\langle\langle {\bf{p.p'} \rangle\rangle}}{p^2}, 
\end{align}
where $p^2=|{\bf p}|^2$ and $A$ is the drag coefficient of the heavy quark. Similarly, the momentum diffusion $B_{ij}$ can be decomposed in terms of longitudinal and transverse components as follows, 
\begin{align}\label{18}
&B_{i,j} = \left(\delta_{ij}-\frac{p_ip_j}{p^2}\right) B_0(p^2)+\frac{p_ip_j}{p^2}B_1(p^2),
\end{align}
with the transverse and longitudinal diffusion coefficients respectively take the forms as,
\begin{align}\label{19}
&B_{0}= \frac{1}{4}\left[\langle\langle p'^{2} \rangle\rangle-\frac{\langle\langle ({\bf{p'.p}})^2\rangle\rangle}{p^2} \right],\\ 
&B_{1}= \frac{1}{2}\left[\frac{\langle\langle ({\bf{p'.p})}^2\rangle\rangle}{p^2} -2\langle\langle ({\bf{p'.p})}\rangle\rangle +p^2 \langle\langle 1 \rangle\rangle\right]\label{19.1}.
 \end{align}
In the current analysis, the effect of anisotropy is entering through the momentum distribution function of the effective degrees of freedom. Incorporating the definition of the distribution function as described in Eq.~(\ref{3}), the thermal average of a function $F(p')$  can be decomposed as follows,
\begin{align}\label{20}
   \langle\langle F({p'})\rangle\rangle=\langle\langle F({p'})\rangle\rangle_0+\langle\langle F({p'})\rangle\rangle_{a},
\end{align}
where the isotropic and anisotropic parts respectively take the forms as follows,
\begin{align}\label{21}
   \langle\langle F({p'})\rangle\rangle_0&=\frac{1}{2E_p}\int{\frac{d^3{\bf q}}{(2\pi)^32E_q}}\int{\frac{d^3{\bf q}'}{(2\pi)^32E_{q'}}}\int{\frac{d^3{\bf p}'}{(2\pi)^32E_{p'}}}\nonumber\\ &\times\frac{1}{\gamma_{HQ}}\sum|\mathcal{M}_{2\rightarrow 2}|^2(2\pi)^4\delta^4 (P+Q-P'-Q') \nonumber\\ &\times f^0_{k}({\bf{q}})\Big(1+a_k f^0_{k}({\bf{q'}})\Big)F({p'}),
\end{align}
and, 
\begin{align}\label{22}
   \langle\langle F({p'})\rangle\rangle&_{a}=\frac{1}{2E_p}\int{\frac{d^3{\bf q}}{(2\pi)^32E_q}}\int{\frac{d^3{\bf q}'}{(2\pi)^32E_{q'}}}\int{\frac{d^3{\bf p}'}{(2\pi)^32E_{p'}}}\nonumber\\ &\times\frac{1}{\gamma_{HQ}}\sum|\mathcal{M}_{2\rightarrow 2}|^2(2\pi)^4\delta^4 (P+Q-P'-Q') \nonumber\\ &\times \Big[\delta f_k({\bf{q}})\Big(1+a_k f^0_{k}({\bf{q'}})\Big) +a_k f^0_{k}({\bf{q}})\delta f_{k}({\bf{q'}})\Big]F({p'}).
\end{align}
Substituting Eq.~(\ref{20}) in  Eq.~(\ref{17})-Eq.~(\ref{19.1}), we obtain the heavy quark transport coefficients due to the elastic collisions in the anisotropic medium as, 
\begin{align}\label{23}
X_c= X_{c\,0}+ X_{c\,a},
\end{align}
where $X_{c\,0}$ is the transport coefficient in the equilibrated medium within the EQPM description. The term $ X_{c\,a}$ denotes the non-equilibrium corrections to the heavy quark transport coefficients due to the anisotropy in the QGP medium. In general, heavy quark transport coefficient due to elastic scattering can be schematically described as,
\begin{align}\label{24}
    X_c=\int \text{phase space}\times \text{ interaction}\times \text{transport part}.
\end{align}
It is important to note that the thermal medium interactions are embedded in the analysis through the effective fugacities. The EQPM description modifies the interaction strength through the effective coupling in the medium while defining the scattering matrix~\cite{Mitra:2016zdw}. The integrals described in Eq.~(\ref{21}) and Eq.~(\ref{22}) can be further simplified and solved in the center-of-momentum frame of the colliding particles and is well investigated, see Refs.~\cite{Svetitsky:1987gq,GolamMustafa:1997id} for detailed discussions.
 \subsection{\bf{ Radiative processes}}
Now, we consider the contribution of the radiative process of the heavy quarks in the medium to the transport coefficients. 
We consider the radiative process  $HQ(P)+l(Q)\rightarrow HQ(P^{'})+l(Q^{'})+g(K_5)$, where $K_5=(E_5, k_{\perp}, k_z)$ is the four-momentum of the emitted gluons. To evaluate the transport coefficients due to the radiative process, the two-body phase space and the matrix element of the elastic collisional process in Eq.~(\ref{24}) need to be replaced with the three body counterparts, and hence we have~\cite{Mazumder:2013oaa},
\begin{align}\label{25}
		&X_{r}=\frac{1}{2E_p}\int{\frac{d^3 {\bf q}}{(2\pi)^32E_q}}\int{\frac{d^3{\bf q}'}{(2\pi)^32E_{q'}}}\int{\frac{d^3{\bf p}'}{(2\pi)^32E_{p'}}}\frac{1}{\gamma_{HQ}}\nonumber\\ &\times \int{\frac{d^3{\bf k}_5}{(2\pi)^32E_5}}\sum{|\mathcal{M}|^2_{2\rightarrow 3}(2\pi)^4\delta^4 }(P+Q-P'-Q'-K_5)\nonumber\\
		&\times f_k({\bf q})(1+a_k f_{k}({\bf q}')(1+ \hat{f}(E_5))\theta_1(\tau-\tau_F)\theta_2(E_p-E_5),
        \end {align}
where $\hat{f}(E_5)=\frac{1}{\exp{(\beta E_5)}-1}$ is the the distribution of the emitted gluon. Here, the theta functions impose restrictions on the gluon radiation. The function $\theta(E_p-E_5)$ put the constraint that the energy of emitted gluon should be less than the energy of heavy quarks. Similarly, the theta function  $\theta_1(\tau-\tau_F)$ denotes that the formation time of the gluon ($\tau_F$) should be less than the scattering time ($\tau$) that accounts for the  Landau-Pomeranchuk-Migdal (LPM) effect~\cite{Gyulassy:1993hr,Klein:1998du}. Note that the current focus is on the soft gluon emission $i.e.$, $K_5\rightarrow$ {0}. The invariant amplitude for radiative processes ($2\rightarrow 3$ process) $|\mathcal{M}|^2_{2\rightarrow 3}$ can be described in terms of $|\mathcal{M}|^2_{2\rightarrow 2}$ for collisional process, the dead cone factor, and the transverse momentum of the emitted gluon ($k_\perp$)  as follows~\cite{Abir:2011jb},
\begin{align}\label{26}
    |\mathcal{M}|^2_{2\rightarrow 3}=|\mathcal{M}|^2_{2\rightarrow 2}\times 12g^2\frac{1}{k^2_\perp}\left(1+\frac{M_{HQ}^2}{s}e^{2\eta}\right)^{-2},
\end{align} 
where $s=(P+Q)^2$ is the Mandelstam variable and $\eta$ is the rapidity of emitted massless gluons. The term $\big(1+\frac{M_{HQ}^2}{s}e^{2\eta}\big)^{-2}$ is the suppression factor due to the dead cone effect~\cite{Dokshitzer:2001zm}. Substituting Eq.~(\ref{3}) in Eq.~(\ref{25}), we obtain the radiative counterpart of the heavy quark transport coefficients as follows, 
\begin{align}\label{27}
X_{r} = X_{r\,0} +  X_{r\,a},
\end{align}
where the equilibrium and anisotropic parts respectively take the following forms,
\begin{widetext}
\begin{align}\label{28}
X_{r\,0}&=\frac{1}{2E_p}\int{\frac{d^3{\bf q}}{(2\pi)^32E_q}}\int{\frac{d^3{\bf q}'}{(2\pi)^32E_{q'}}}\int{\frac{d^3{\bf p}'}{(2\pi)^32E_{p'}}} \int{\frac{d^3{\bf k}_5}{(2\pi)^32E_5}}\frac{1}{\gamma_{HQ}}\sum{|\mathcal{M}|^2_{2\rightarrow 3}(2\pi)^4\delta^4 }(P+Q-P'-Q'-K_5)\nonumber\\ 
&\times f_k^0({\bf q})(1+a_k f^0_{k}({\bf q}')\Big(1+ \hat{f}(E_5)\Big)\theta_1(\tau-\tau_F)\theta_2(E_p-E_5),
\end {align}
\begin{align}\label{29}
 X_{r\,a}&=\frac{1}{2E_p}\int{\frac{d^3{\bf q}}{(2\pi)^32E_q}}\int{\frac{d^3{\bf q}'}{(2\pi)^32E_{q'}}}\int{\frac{d^3{\bf p}'}{(2\pi)^32E_{p'}}} \int{\frac{d^3{\bf k}_5}{(2\pi)^32E_5}}\frac{1}{\gamma_{HQ}}\sum{|\mathcal{M}|^2_{2\rightarrow 3}(2\pi)^4\delta^4 }(P+Q-P'-Q'-K_5)\nonumber\\
&\times \Big[\delta f_k({\bf{q}})\Big(1+a_k f^0_{k}({\bf{q'}})\Big) +a_k f^0_{k}({\bf{q}})\delta f_{k}({\bf{q'}})\Big]\Big(1+ \hat{f}(E_5)\Big)\theta_1(\tau-\tau_F)\theta_2(E_p-E_5).
\end {align}
\end{widetext}
 \begin{figure*}
 \centering
 \subfloat{\includegraphics[scale=0.39]{drLHCP.eps}}
 \hspace{1 cm}
 \subfloat{\includegraphics[scale=0.39]{diLHCP.eps}}
  \hspace{1 cm}
 \subfloat{\includegraphics[scale=0.39]{dilLHCP.eps}}
\caption{Momentum dependence of heavy quark drag coefficient (left panel), diffusion coefficient $B_0$ (middle panel), and $B_1$ (right panel) for the LHC energy at $T=480$ MeV in an anisotropic QGP medium.}
\label{f2}
\end{figure*}
Substituting Eq.~(\ref{26}) in Eq.~(\ref{25}), we can represent the radiative counterpart of the heavy quark transport coefficients in terms of the collisional part as follows,
\begin{align}\label{30}
    X_{r}=&X_{c}\int\frac{d^3{\bf k}_5}{(2\pi)^32E_5}12g^2\frac{1}{k^2_\perp}\left(1+\frac{M_{HQ}^2}{s}e^{2\eta}\right)^{-2}\nonumber\\
    &\times\Big(1+ \hat{f}(E_5)\Big)\theta_1(\tau-\tau_F)\theta_2(E_p-E_5). 
\end{align}
The Eq.~(\ref{30}) can be further simplified by converting the massless gluon four-momentum in terms of rapidity variable, and we have 
\begin{align}\label{31}
    &E_5=k_\perp\cosh \eta, &&k_z=k_\perp\sinh \eta, 
\end{align}
with $d^3{ k}_5=d^2{ k}_\perp dk_z=2\pi k_\perp^2 dk_\perp \cosh\eta d\eta$. The interaction time is related to the interaction rate $\Lambda$ and the function $\theta_1(\tau-\tau_F)$ put the constraint as~\cite{Das:2010tj},
\begin{align}\label{32}
    &\tau=\Lambda^{-1}>\tau_F=\frac{\cosh\eta}{k_\perp},
\end{align}
which indicates $k_\perp>\Lambda\cosh\eta=(k_\perp)_{\text{min}}$, where $(k_\perp)_{\text{min}}$ denotes the minimum value of $k_{\perp}$. Further, from function $\theta_2(E_p-E_5)$, we have,
\begin{align}\label{33}
   &E_p>E_5=k_\perp\cosh\eta,
  &&(k_\perp)_{\text{max}}=\frac{E_p}{\cosh\eta}.
\end{align}
For the case of soft gluon emission of the heavy quarks in the medium, we have $E_5=k_\perp \cosh \eta \ll T$ such that the distribution function of the emitted massless gluons can be approximated as,	
\begin{align}\label{34}
 1+\hat{f}(E_5)=1+\frac{T}{k_\perp\cosh \eta}\approx\frac{T}{k_\perp\cosh \eta}.   
\end{align}
The equilibrium and anisotropic parts of the heavy quark transport coefficients can be obtained by solving Eq.~(\ref{28}) and Eq.~(\ref{29}) within the above approximations. The effective drag and diffusion coefficients of the heavy quarks due to the collisional and radiative processes in the anisotropic medium can be obtained by adding collisional and radiational parts assuming that the elastic collision and soft gluon emission take place independently in the QGP. Hence, from Eq.~(\ref{23}) and Eq.~(\ref{27}), the effective transport coefficient within the EQPM takes the form as follows,
\begin{align}\label{36}
	X=X_{0}+ X_a,
\end{align}
where $X_0$ is the net equilibrium part and $X_a$ is the total anisotropic contributions to the radiative and collisional parts. Hence, we have
\begin{align}\label{36.1}
	&X_0=X_{c\,0}+X_{r\,0}, && X_a=  X_{c\,a}  + X_{r\,a}.
\end{align}
We shall now proceed to investigate the effect of thermal medium interactions and anisotropy of the medium to the heavy quark transport coefficients and energy loss while including the collisional and radiative  processes.
 \begin{figure*}
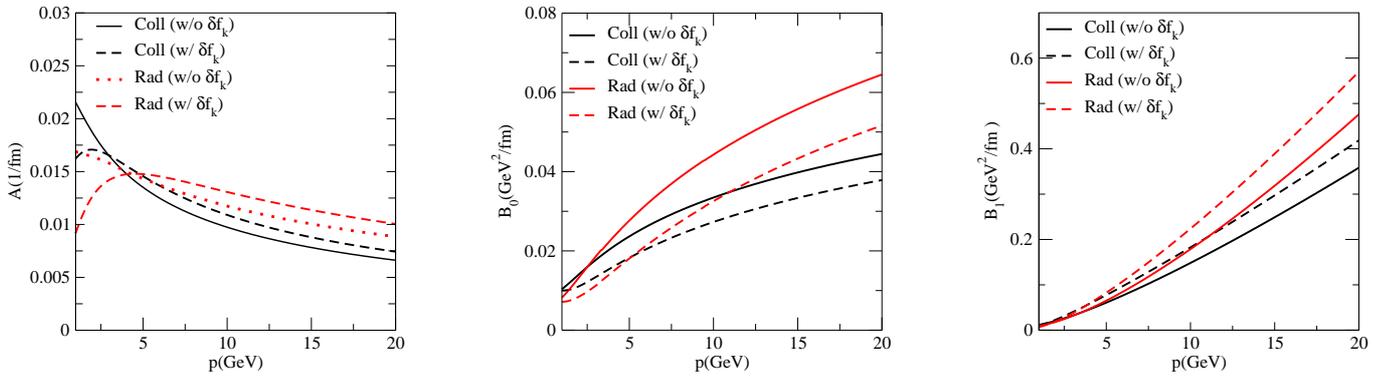

 \centering
 \subfloat{\includegraphics[scale=0.39]{drRHICP.eps}}
 \hspace{1 cm}
 \subfloat{\includegraphics[scale=0.39]{diRHICP.eps}}
  \hspace{1 cm}
 \subfloat{\includegraphics[scale=0.39]{dilRHICP.eps}}
\caption{Momentum dependence of heavy quark drag coefficient (left panel), diffusion coefficient $B_0$ (middle panel), and $B_1$ (right panel) for the RHIC energy at $T=360$ MeV in an anisotropic QGP medium.}
\label{f2.1}
\end{figure*}
 \section{Results and Discussions}
\subsection{Heavy quark radiative and collisional process in the anisotropic medium}
We initiate the discussions with the momentum dependence of heavy quark drag and diffusion coefficients in the equilibrated medium while incorporating the thermal medium interactions. The EQPM description of the heavy quark transport coefficients is described in Eq.~(\ref{36}). The drag and diffusion coefficients of heavy quarks in the non-interacting  QGP (the medium described by an ultra-relativistic gas of quarks/antiquarks and gluons (ideal EOS))  has been initially studied by including collisional interactions~\cite{Svetitsky:1987gq} and later with the radiative process in the medium~\cite{Mazumder:2013oaa}. In the current analysis, the EoS effects are incorporated through the momentum distribution function of the effective degrees of freedom via effective fugacities and through the effective coupling. The momentum behaviour of the drag and diffusion coefficients is depicted in Fig.~\ref{f2} for the LHC energy and in Fig.~\ref{f2.1} for the RHIC energy. The effect of thermal medium interactions on the heavy quark transport coefficients has already been studied in Ref.~\cite{Das:2012ck}. In the asymptotic limit, the EQPM results reduce back to the results of non-interacting QGP. The impact of the gluon radiation by heavy quark on the drag and diffusion coefficients are prominent throughout the chosen range of momentum for the RHIC and LHC energies. This indicates that the inclusion of the radiative counterpart is essential for the analysis of measured observables from collision experiments at the RHIC and LHC. It is observed that the collisional part exceeds the radiative part within the EQPM for low heavy quark momentum, especially at the higher temperature regimes. However, at higher momentum regimes, the radiative contribution to the transport coefficients is dominant over the collisional counterparts.  

We have incorporated the effects of anisotropy to the heavy quark transport through the momentum distribution functions of the constituent particles in the medium. The heavy quark collisional and radiative processes are sensitive to the anisotropy in the medium, and the effect can be quantified in terms of the drag and diffusion coefficients. The momentum dependence of drag and diffusion coefficients is depicted in an anisotropic medium in Fig.~\ref{f2} for the LHC energy at $T=480$ MeV and in Fig.~\ref{f2.1} for the RHIC energy at $T=360$ MeV. The effect of instability and hence the anisotropy in the medium is related  with the phenomenologically known jet quenching parameter in the collision experiments. For the quantitative estimation, we choose $\hat{q}_k=3.7 T^3$~\cite{Burke:2013yra}, the thermalization time $\tau= 0.6$ fm for the LHC energy, and $\hat{q}_k=4.6 T^3$~\cite{Burke:2013yra}, $\tau= 0.9$ fm for the RHIC energy in the current analysis. The impact of anisotropy on the collisional process has been investigated on the Ref.~\cite{Chandra:2015gma}. It is observed that the radiative process of the heavy quark significantly modifies the transport coefficients in the anisotropic medium. The impact of anisotropy on the drag coefficient due to the elastic and inelastic processes for the LHC energy is shown in Fig.~\ref{f2} (left panel). We observed that the instability creates a lesser hindrance for the heavy quark motion in the QGP while emitting soft gluon radiation at low momentum. However, the collisional and radiational contributions of the drag coefficient increases with the anisotropy for the heavy quark momentum above $p=5$ GeV at the LHC energies. Notably, the effect of anisotropy is more prominent in the radiational process in comparison with the collisional process of heavy quarks in the medium. Further, we verified that the same observation on the effect of anisotropy on the heavy quark coefficient holds true for the RHIC energies too. 

The heavy quark diffusion coefficients $B_0$ and $B_1$ are plotted as a function of momentum in Fig.~\ref{f2} (middle and right panels) for LHC energy and in Fig.~\ref{f2.1} (middle and right panels) for the RHIC energy. The radiative process is seen to have a dominant contribution to the heavy quark momentum diffusion in comparison with the collisional process in the medium, especially for the high momentum regimes. In contrast to the drag coefficient, the anisotropy has a weaker dependence on $B_0$ and $B_1$ for the low momentum regime. As expected, the momentum anisotropy of the medium has different ramifications on the heavy quark transport in various directions. This is reflected in the momentum dependence of $B_0$ and $B_1$ in the anisotropic medium. It is observed that the anisotropy in the medium suppresses the diffusion coefficient $B_0$, whereas $B_1$ shows the opposite behaviour in the anisotropic medium.
\begin{figure*}
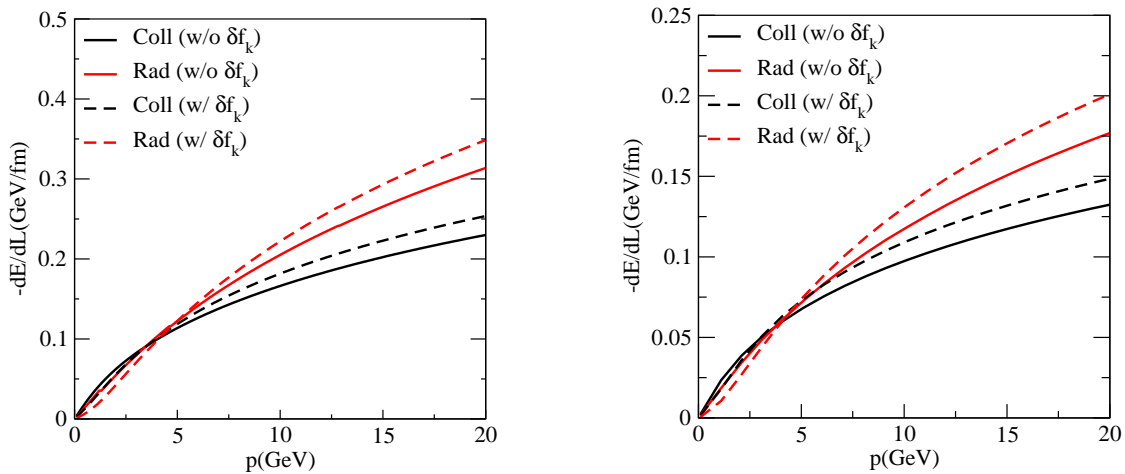

 \centering
 \subfloat{\includegraphics[scale=0.5]{elhc1.eps}}
 \hspace{1.5 cm}
 \subfloat{\includegraphics[scale=0.5]{erhic1.eps}}
\caption{Momentum dependence of collisional and radiative energy loss of the heavy quark in the medium for the LHC energy at $T=480$ MeV (left panel) and the RHIC energy at $T=360$ MeV (right panel).}
\label{f3}
\end{figure*}     
\subsection{Heavy quark collisional and radiative  energy loss}

Heavy quark travels through the medium and may lose its energy by elastic collisions with the constituent particles in the medium and by radiating gluons in the medium. The net energy loss of the heavy quark in the medium can be 
quantified in terms of the drag force that offers the resistance to the heavy quark motion. The differential heavy quark energy loss can be defined as follows~\cite{GolamMustafa:1997id},
\begin{equation}\label{37}
-\dfrac{dE}{dL}=A(p^2, T)p.
\end{equation}
The energy loss of heavy quark due to the hard and soft collision processes in the medium has been investigated in Ref.~\cite{Braaten:1991we}. Further, the additional mechanism of soft gluon radiation enhances the energy loss. In Fig.~\ref{f3}, the heavy quark energy loss in the medium due to elastic and inelastic processes are plotted as a function of its momentum for the LHC and RHIC energies. The realistic EoS effects suppress the collisional and radiative energy loss of the heavy quark in the interacting medium.  Notably, the EoS effects to the heavy quark energy loss will be negligible in the asymptotic limit of the temperature as the medium behaves as an ultra-relativistic non-interacting system at very high temperatures. The heavy quark energy loss in the QGP medium critically depends on its momentum and temperature of the medium. In Ref.~\cite{Kurian:2020orp}, the authors have reported that the non-equilibrium corrections such as shear and bulk viscous corrections have a weaker dependence on the heavy quark energy loss in the QGP medium. In the current analysis, we studied the effect of the momentum anisotropy induces from the instabilities in the non-equilibrium QGP. It is observed that the momentum anisotropy of the medium has a visible impact on the heavy quark energy loss. This observation holds true for both the LHC and RHIC energies.  The collisional and radiational energy loss of heavy quark for the RHIC is reduced by approximately 30\% in comparison with that for the LHC energy. It is also important to emphasize that the radiative contribution to the energy loss dominates the collisional part above $p=5$ GeV in the isotropic and anisotropic QGP medium. However, in the low momentum regions, the collisional energy loss is higher than that from the soft gluon radiation of heavy quark in the medium for both the LHC and RHIC energies.   
\begin{figure*}
 \centering
 \subfloat{\includegraphics[scale=0.305]{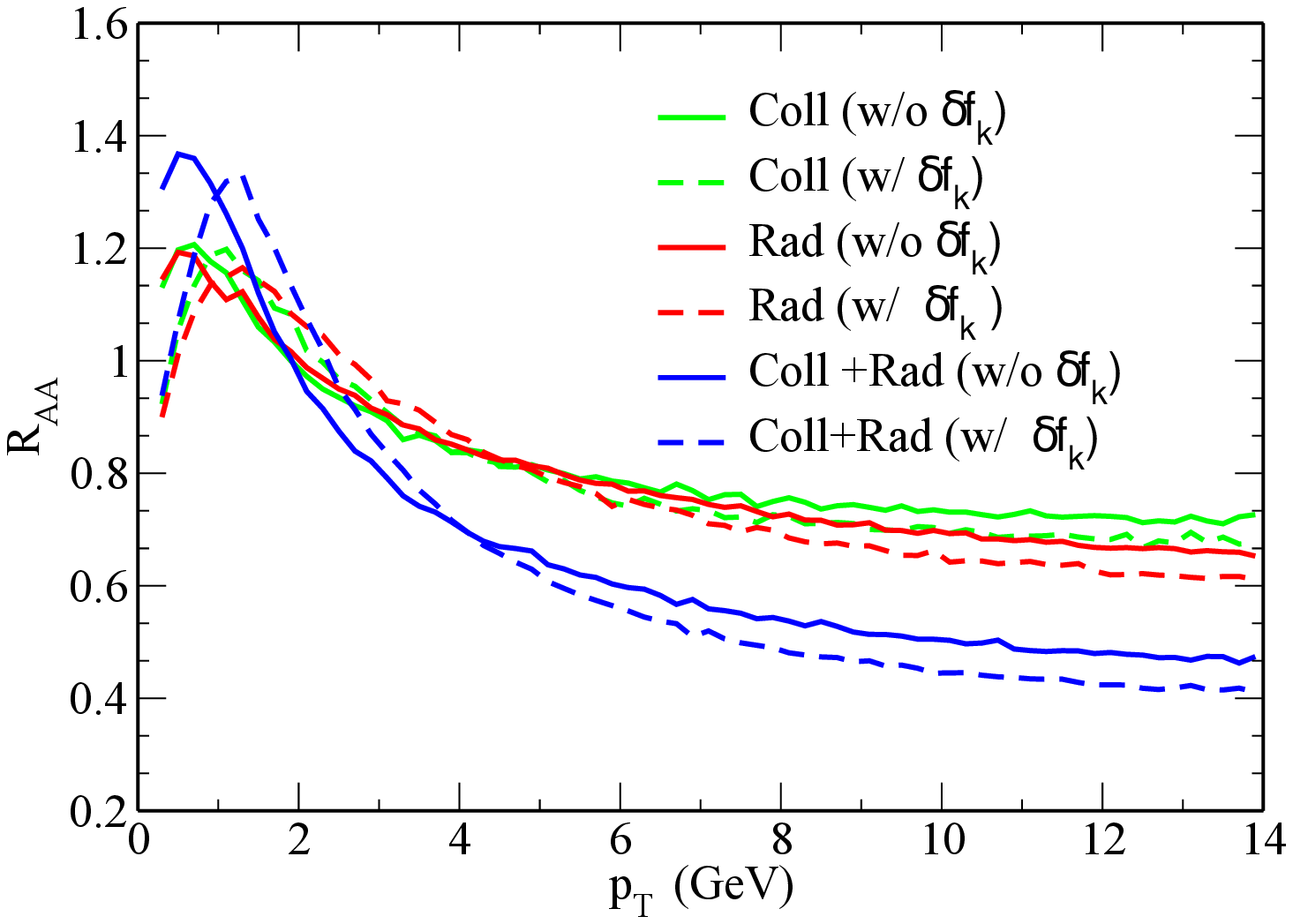}}
 \subfloat{\includegraphics[scale=0.305]{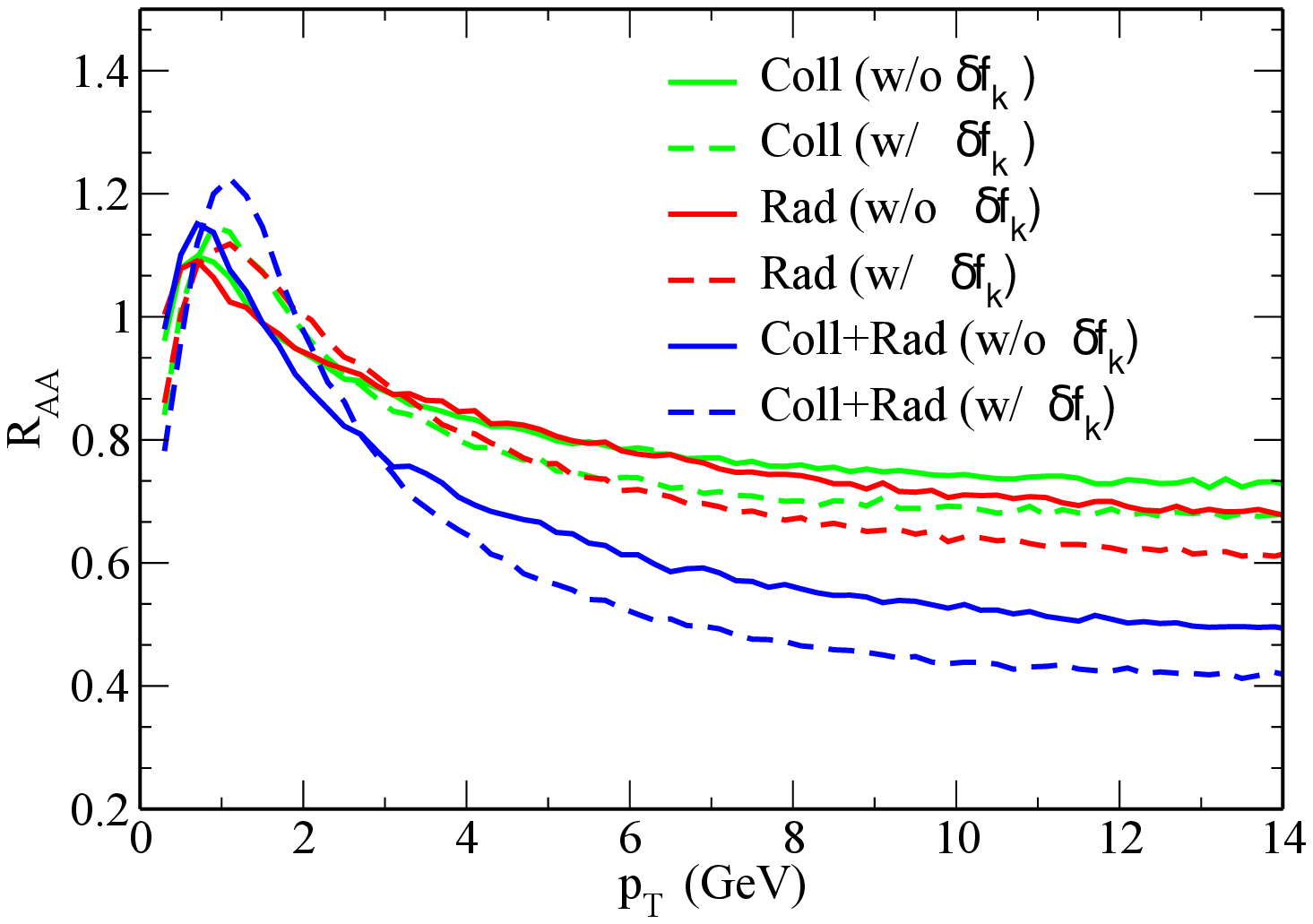}}
\caption{The nuclear suppression factor ($R_{AA}$) as a function of $p_T$ for charm quark for the LHC energy at $T=480$ MeV (left panel) and the RHIC energy at $T=360$ MeV (right panel).}
\label{f4}
\end{figure*}     
\subsection{Nuclear Modification factor $R_{AA}$}
To study the impact of anisotropy  by heavy quarks in the medium on the experimental observable, we have estimated the nuclear suppression factor $R_{AA}$, employing the charm quark distribution functions at initial time $t=\tau_i$ and  final time $t=\tau_f$ as $R_{AA}=\frac{f_{\tau_f} (p )}{f_{\tau_i} (p)}$. This requires the adequate knowledge of heavy quark dynamics in the QGP medium. The standard approach to obtain the heavy quark momentum evolution in the medium is to solve the Fokker-Plank equation stochastically by the Langevin equations.  The Langevin equations of motion for heavy quarks take the following forms~\cite{Moore:2004tg,Das:2013kea},
\begin{align}
   &dx_i=\frac{p_i}{E} dt,\\
   & dp_i=-Ap_i\, dt+C_{ij}\rho_j\sqrt{dt},
\end{align}
where $dx_i$ and $dp_i$ are respectively the position and momentum shift in each time interval $dt$. Here, $A$ is the drag force and $C_{ij}$ denotes the covariance matrix that describes stochastic force in terms of independent Gaussian-normal distributed random variable $\rho_j$ with $\langle\rho_i\rho_j\rangle=\delta_{ij}$ and $\langle\rho_i\rangle=0$. The matrix $C_{ij}$ is related to the heavy quark momentum diffusion coefficients as follows,
\begin{align}
C_{ij} = \sqrt{2B_0}\left(\delta_{ij}-\frac{p_ip_j}{p^2}\right)+\sqrt{2B_1}\frac{p_ip_j}{p^2}B_1.
\end{align}
In the limit $B_0=B_1=D$, we have $C_{ij}=\sqrt{2D}\delta_{ij}$. It is important to note that this assumption is strictly valid for the static limit $(p\rightarrow 0)$, and is also usually employed at finite momentum for heavy quark motion in the QGP medium~\cite{Rapp:2018qla,Moore:2004tg,vanHees:2005wb,vanHees:2007me,Scardina:2017ipo,Cao:2015hia}. In the momentum space, at $\tau_i$,  the charm quarks are distributed according to the Fixed Order + Next-to-Leading Log (FONLL) calculations, taken from Refs.\cite{Cacciari:2005rk,Cacciari:2012ny}. Our aim is to highlight the impact of the anisotropy presented in this manuscript on $R_{AA}$. We have computed the $R_{AA}$ in a static medium at a fixed temperature for both the LHC and RHIC energies at the level of charm quarks. To study the heavy quark momentum evolution within Langevin dynamics, we implemented both the drag and diffusion coefficients presented in Fig.~\ref{f2} and Fig.~\ref{f2.1}.
In this present calculation we consider $\tau_f$=6 fm/c which can be roughly taken as the typical lifetime of  QGP produced at RHIC and LHC energies. In a future effort, we will study the heavy quarks observable in an expanding medium, including hadronization mechanics and the possible impact of  fluctuation-dissipation theorem.

The momentum dependence of $R_{AA}$ is estimated in the anisotropic medium while including the collisional and radiative processes of heavy quark in the QGP within the Langevin dynamics. The effects of anisotropy and soft gluon emission of heavy quarks in the medium are entering through the drag and diffusion coefficients, adding both collisional and radiative drag and diffusion coefficients~\cite{Das:2010tj,Mazumder:2011nj,Gossiaux:2006yu}. An alternative approach to include the radiative energy loss into the Langevin framework can be found in Refs.~\cite{Cao:2013ita,Cao:2015hia}.

In Fig.~\ref{f4}, $R_{AA}$ is plotted as a function of $p_T$ at the LHC (left panel) and RHIC energies (right panel). The gluon emission by heavy quarks in the QGP medium substantially modifies the nuclear suppression factor. We observe a strong suppression (small $R_{AA}$) with the inclusion of the radiative process along with the elastic collisional process of heavy quarks in the QGP medium. Notably, in the high $p_T$ regimes, the $R_{AA}$ due to the radiative process is smaller than that due to the elastic interaction. We observe a similar trend for $R_{AA}$ for both the LHC and RHIC energies.  The impact of anisotropy of the medium to the $R_{AA}$ is further displayed in Fig.~\ref{f4}. The effects of momentum anisotropy on the heavy quark transport coefficients seem to have a visible impact on the $R_{AA}$ for the LHC as well as RHIC energies. In the higher $p_T$ regimes, the anisotropic effects decrease the $R_{AA}$ leading to a stronger suppression. This observation is a consequence of the fact that the instability (momentum anisotropy) offers larger hindrance for the charm quark motion in the QGP at high momentum.

  \section{Summary and Outlook}
In conclusion, we have investigated the dynamics of heavy quarks undergoing radiative energy loss along with the elastic collisions with the constituent particles in an anisotropic hot QCD medium. The elastic and inelastic (soft gluon radiation) interactions of the heavy quark with the medium have been  studied in terms of drag and diffusion coefficients within the framework of the Fokker-Planck approach. The thermal QCD medium interactions are incorporated in the analysis through the temperature-dependent quark, antiquark, and gluonic effective fugacities within the EQPM description. We have observed that the gluon radiation emission has a significant contribution to the heavy quark drag and diffusion coefficients in the QGP medium. 

We have conducted a systematic analysis on the momentum dependence of the heavy quark in an anisotropic (momentum) hot QCD  medium. The momentum anisotropy that may lead to Chromo-Weibel instability leads to an effective Boltzmann-Vlasov equation. The non-equilibrium momentum distribution of the effective degrees of freedom is obtained by solving the ensemble-averaged diffusive Vlasov-Boltzmann equation. The effects of momentum anisotropy on the heavy quark transport coefficients are seen to be quite significant for both the collisional and radiative processes in the medium. Moreover, these anisotropic corrections induced from the instabilities in the QCD medium are essential to maintain theoretical consistency in the description of heavy quark transport in the near-equilibrium medium. Further, we have studied the impact of momentum anisotropy for the collisional and radiative energy losses of the heavy quark in the hot QCD medium. The effects of anisotropy and gluon emission by heavy quarks in the medium are found to have noticeable effects in the momentum dependence of the nuclear suppression factor $R_{AA}$ both at the RHIC and LHC energies.

The effects of realistic EoS and momentum anisotropy to the heavy quark transport coefficients may have a significant impact on flow coefficients of heavy mesons at the LHC and RHIC. We intend to investigate these phenomenological aspects in the near future. The drag and diffusion coefficients, while including the radiative and collisional processes of the heavy quark in the expanding medium within the framework of dissipative hydrodynamics, will be a timely work to follow. The gluon emission by the charm quark in a magnetized medium is another direction to work in the near future.

\section*{acknowledgments}
M.K. would like to acknowledge Indian Institute of Technology Gandhinagar for Institute postdoctoral fellowship. We are thankful to Jan-e-Alam  and Trambak Bhattacharyya for useful discussions. This work is conducted under the SERB (Science and Engineering Research Board) Core Research Grant: CRG/2020/002320. We record our deep sense of gratitude to the people of India for their generous support for the investigations in fundamental sciences. 

\bibliography{ref}{}

\end{document}